# Symmetry, incommensurate magnetism and ferroelectricity: the case of the rare-earth manganites $RMnO_3$

J L Ribeiro

Departamento de Física, Universidade do Minho, 4710-057 Braga, Portugal

jlr@fisica.uminho.pt

**Abstract**. The complete irreducible co-representations of the paramagnetic space group provide a simple and direct path to explore the symmetry restrictions of magnetically driven ferroelectricity. The method consists of a straightforward generalization of the method commonly used in the case of displacive modulated systems and allows us to determine, in a simple manner, the full magnetic symmetry of a given phase originated from a given magnetic order parameter. The potential ferroic and magneto-electric properties of that phase can then be established and the exact Landau free energy expansions can be derived from general symmetry considerations.

In this work, this method is applied to the case of the orthorhombic rare-earth manganites $RMnO_3$. This example will allow us to stress some specific points, such as the differences between commensurate or incommensurate magnetic phases regarding the ferroic and magnetoelectric properties, the possible stabilization of ferroelectricity by a single irreducible order parameter or the possible onset of a polarization oriented parallel to the magnetic modulation. The specific example of $TbMnO_3$ will be considered in more detail, in order to characterize the role played by the magneto-electric effect in the mechanism for the polarization rotation induced by an external magnetic field.

## 1. Introduction

In the last few years, there has been a great deal of interest in several metal compounds, where ferroelectricity is induced by a transition to a complex magnetic state. Examples of this class of materials are systems like $RMnO_3$ (R=Gd, Dy or Tb), $RMn_2O_5$ (R=Tb or Y), $MnWO_4$, $NiVa_2O_8$, $CoCr_2O_4$, $CuFeO_2$, $CuO$ or some complex hexagonal ferrites like $(Ba,Sr)_2Zn_2Fe_{12}O_{22}$ [1-9]. In addition, some of these systems show remarkable effects such as the ability of a magnetic field to rotate or stabilize an electrical polarization, as in the case of $TbMnO_3$ or $GdMnO_3$, respectively [10-12].

In this class of materials, and in contrast with conventional multiferroics like $BiFeO_3$ or $BiMnO_3$, the paramagnetic phase is also paraelectric and the ferroelectric polarization is driven by a modulated magnetic order parameter. This improper ferroelectricity reveals therefore a close resemblance with that originated from lattice modulations, as in the case of systems like $RbZn_2Cl_4$, $(C(NH_3)_4)_2CoCl_4$ or $(CH_3)_3NCH_2COO.CaCl_2.2H_2O$ [13-15].

The first efforts to understand this magnetically driven ferroelectricity have been mainly based on the search for microscopic mechanisms coupling magnetic moments and lattice displacements, such as the symmetric superexchange striction or the anti-symmetric exchange (Dzyaloshinskii-Moryia mechanism [16-17]). Also, phenomenological approaches like the so called "spin current" [18] or the "spiral formulation"[19] have been explored. Here, a spiral magnetic structure with a rotation axis

$\vec{e}$ and modulation wavevector $\vec{k}$ breaks inversion and gives rise to a polarization $\vec{P} // \vec{e} \times \vec{k}$. However, a satisfactory and integrated description of the effect has not yet been achieved.

Establishing the symmetry constrains upon the onset of a polarization vector, toroidal moment or lattice distortion as a result of the stabilization of a given magnetic structure, independently of the microscopic mechanisms involved, is a valuable contribution to further understand the phenomena. Symmetry imposes a rigid framework for the possible interactions between the different degrees of freedom and guarantees the overall consistency of the observed laws. Moreover, given the direct connection between symmetry and phenomenology, via Landau theory, this represents a first step to establish more detailed models and to further elucidate the relevant microscopic mechanisms involved.

The analysis of the symmetry requirements for a magnetically driven ferroelectricity has been pursued in a number of studies which have followed different approaches [20-23]. In this work, the method adopted uses the complete irreducible co-representations (CICR´s) of the magnetic space group $G$ of the parent structure, in order to determine the full symmetry of a given modulated magnetic phase. The specific example of the orthorhombic rare-earth manganites $RMnO_3$ will allow us to stress some aspects related to the possible ferroic properties of these compounds, emphasizing points such as the role of the magneto-elastic coupling, the essential differences between commensurate and incommensurate modulations regarding the ferroic and magneto-electric properties or the possible onset of an improper polarization due to one irreducible magnetic order parameter. The specific case of $TbMnO_3$ will be further considered, in order to establish the role of the magneto-electric effect in the mechanism for the polarization rotation induced by an external magnetic field.

## 2. The method for the symmetry analysis: the case of the $RMnO_3$ compounds

The approach of working with the CICR`s allows us to use full image of $G$ in the order parameter space instead of considering only the unitary small representation of the vector $\vec{k}$, possibly complemented by the *ad-hoc* inclusion of spatial inversion [20]. This has a number of advantages. In general, this method incorporates in a natural way the physical representations related to $\vec{k}$ and $-\vec{k}$, even if inversion is not a symmetry operation of $G$. Also, by integrating the full magnetic symmetry, this method allows us to analyze the possible stabilization of the different ferroic parameters, which are precisely distinguished by their different behaviour under spatial inversion and time reversal. Finally, the use of the CICR`s allows us to deal with commensurate and incommensurate phases in an unified way, to stress their differences regarding the possible magneto-electric coupling and to use the concept of superspace groups in the case of a magnetic incommensurate order parameter.

Let us consider the example of the orthorhombic rare-earth manganites $RMnO_3$. Here, $G$ is the paramagnetic group $(Pnma)´$, $\vec{k}_1 = \delta(T)\vec{a}^*$ (which corresponds to the $\Sigma$ line in the reciprocal cell), the star is $\{\vec{k}_1^*\} = \{\vec{k}_1 = \vec{k}; \vec{k}_2 = -\vec{k}\}$ and the small co-representations are one-dimensional. The magnetic modulation in a given ordered phase will be described by the field $\vec{S}(\vec{T}, \nu) = \sum_{n,j} S_{nj} \vec{e}_\nu e^{i\vec{k}_n \bullet \vec{T}}$, where $\vec{e}_\nu(\vec{k}_n, j)$ are the eigenvectors of the mode, $\nu$ labels the spins in the reference unit cell and $\vec{T}$ denotes a given unit cell of the original structure. The field $\vec{S}$ represents the magnetic order parameter (OP).

The complete co-representation of $G$ in the OP space will be formed by the set of matrices $D_{op}(\vec{k}_1, \{R; \vec{t}\})$ representing the symmetry operations $\{R; \vec{t}\} \subset G$ in the vector space generated by the components $S_{nj}$. Then, the normal co-ordinates of the active mode will be transformed, under the action of $\{R; \vec{t}\}$, as $S'_{mi} = [D_{op}(\vec{k}_1, \{R; \vec{t}\})]_{mnij} S_{nj}$. This means that the knowledge of the matrices $D_{op}(\vec{k}_1, \{R; \vec{t}\})$ will allow us to characterize how the order parameter is transformed under the different symmetry operations of *G* and to establish, consequently, the symmetry of the modulated phase.

For the RMnO$_3$ compounds, and in the case of one irreducible OP, the field $\vec{S}(\vec{T},\nu)$ can be simply written as $\vec{S}(\vec{x}) = \vec{S}e^{i\vec{k}\cdot\vec{x}} + \vec{S}^*e^{-i\vec{k}\cdot\vec{x}}$. Here, $\vec{S} = \vec{S}_0 e^{i\Phi}$ is the eigenvector of the mode and $\Phi$ its phase, defined with respect to the underlying discrete lattice. Note that, $\vec{S}(\vec{x})$ is real ($\vec{S}_{\vec{k}}^* = \vec{S}_{-\vec{k}}$) and, given its magnetic nature, odd under time reversal $\theta\vec{S}(\vec{x}) = -\vec{S}(\vec{x})$.

There are 12 irreducible magnetic eigenvectors generated by the four *Mn* spins per unit cell. If one labels the spins located at (0,0,½), (½,0,0), (0,½, ½) and (½, ½, 0) as $\vec{S}_1, \vec{S}_2, \vec{S}_3$ and $\vec{S}_4$, respectively, then these eigenvectors $\vec{S}$ correspond to the components of the modes F($\vec{S}_1 + \vec{S}_2 + \vec{S}_3 + \vec{S}_4$), G($\vec{S}_1 - \vec{S}_2 - \vec{S}_3 + \vec{S}_4$), C($\vec{S}_1 - \vec{S}_2 + \vec{S}_3 - \vec{S}_4$) and A($\vec{S}_1 + \vec{S}_2 - \vec{S}_3 - \vec{S}_4$). Each of these eigenvectors is transformed according to one of the four CICR`s of *G* at $\vec{k} = \delta\vec{a}^*$: $(G_x, C_y, A_z \to \Delta_1(A_g))$, $(G_z, F_y, A_x \to \Delta_2(B_{2g}))$, $(F_x, C_z, A_y \to \Delta_3(B_{3g}))$ and $(G_y, C_x, F_z \to \Delta_4(B_{1g}))$. Therefore, the set of matrices $\{\hat{D}(g)\}$ specifying how the symmetry operations of *G* act on the vector space of a particular irreducible OP will correspond to a particular set of CICR matrices. These sets can be readily obtained by following standard methods [24]. For the case of the magnetic space group (*Pnma*)´, the CICR matrices are listed, for example, in the table 1 of [23]. For a reducible OP, the $D_{op}(\vec{k}_1, \{R;\vec{t}\})$ matrices can be constructed from appropriate direct sums of the CICR matrices.

If the modulation wavevector $\vec{k}_1$ is commensurate, the magnetic space group of the ordered phase G´ will be formed by the set of unitary and anti-unitary operations that leave the OP invariant. That is, if $\hat{D}(g)$ and $\hat{D}(\theta g)$ are respectively the matrices induced by the unitary and the anti-unitary operations $\{g;\vec{t}\}$ and $\{\theta g;\vec{t}\}$ of *G* in the vector space generated by the components of the OP, and if the conditions $\hat{T}\times\hat{D}(g)\times\vec{S} = \vec{S}$ or $\hat{T}\times\hat{D}(\theta g)\times\vec{S}^* = \vec{S}$ hold, then $\{g;\vec{t}+\vec{T}\}$ or $\{\theta g;\vec{t}+\vec{T}\}$ will belong to *G*´.

If $\vec{k}_1$ is incommensurate, the symmetry of the ordered phase will be described by a magnetic superspace group. The reasoning leading to the notion of a superspace group can be entirely based on the invariance properties of the Landau free energy and it is essentially independent of the fact that the order parameter is of displacive or magnetic nature [25]. Evidently, if $\vec{k}_1$ is incommensurate, translational symmetry is lost along certain spatial directions. Consequently, the symmetry of the incommensurate phase cannot be described by one magnetic space group. However, the incommensurability of the structure in the ordered phase allows us to shift the phase of the order parameter without changing the free energy of the system. If a certain set of phase translations $\{\vec{\tau}\}$ can be combined with a set $\{g;\vec{t}\}$ of symmetry operations of G, in such a way that the generalized operations $\{g;\vec{t},\tau\}$ keep the free energy and the OP invariant and, in addition, obey the product rule $\{R_2;\vec{t}_2,\vec{\tau}_2\}\times\{R_1;\vec{t}_1,\vec{\tau}_1\} = \{R_2R_1;\vec{t}_2 + R_2\vec{t}_1, \vec{\tau}_2 + \Re(R_2)\vec{\tau}_1\}$, then these operations can be seen as symmetry elements of the incommensurate structure. The matrix $\Re$ given above is defined by the action of *R* on the wavevectors of the star $\{\vec{k}_1^*\}$ according to the procedure described in [25].

Similar to the case of a commensurate wavevector, the incommensurate order parameter will be invariant under the action of the unitary $\{g;\vec{t},\vec{\tau}(g)\}$ or the anti-unitary $\{\theta g;\vec{t},\vec{\tau}(\theta g)\}$ operations if $\hat{\Gamma}(\vec{\tau})\times\hat{D}(g)\times\vec{S} = \vec{S}$ or $\hat{\Gamma}(\vec{\tau})\times\hat{D}(\theta g)\times\vec{S}^* = \vec{S}$. Here, $\hat{\Gamma}(\vec{\tau})$ represents the matrix that expresses, in the space of the order parameter, the phase translation of the magnetic modulation. The set of the symmetry operations that verify these conditions and obey the product rule given above, form the magnetic superspace group of the incommensurate phase.

### 3. Some results for the RMnO$_3$ compounds

A detailed analysis of the symmetry, ferroic properties and free energy expansions of the magnetically modulated phases allowed in the case of the rare-earth manganites will not be considered here. In the following, we will focus on the results concerning some specific points.

*3.1. Ferroelectricity can be induced by an irreducible and commensurate magnetic order parameter.*
The application of the method outlined above to the case of one irreducible and incommensurate (INC) order parameter shows that all the symmetry operations of $G$ give rise to generalized symmetry operations of the incommensurate phase. There is, in this case, a one-to-one relationship between CICR`s and magnetic superspace groups: $\Delta_1 \rightarrow P_a(P_{\bar{1}11}^{nma})$, $\Delta_2 \rightarrow P_a(P_{\bar{1}1S}^{nma})$, $\Delta_3 \rightarrow P_a(P_{\bar{1}SS}^{nma})$ and $\Delta_4 \rightarrow P_a(P_{\bar{1}S1}^{nma})$ [26]. In particular, spatial inversion and time reversal are kept as symmetry operations of the incommensurate structure, when complemented by a phase shift of the modulation corresponding to $\Phi$ or $\pi$, respectively ($\{i;000,\Phi\},\{\theta;000,1/2\}$). This means that, in the case of the RMnO$_3$ compounds, one irreducible and incommensurate phase can never give rise to a polarization $P$, a magnetization $M$ or a toroidal moment $T$. Also, the linear magneto-electric tensor must be null. Homogeneous lattice deformations of symmetry $A_g$ are allowed, together with the lattice modulations $e_{11}(2n\vec{k})$, $e_{22}(2n\vec{k})$ or $e_{33}(2n\vec{k})$, with $n$ integer. Among these, the more important will be those with $n=1$, which correspond to the elastic modulations with one-half of the wavelength of the magnetic modulation detected experimentally. These lattice modulations are secondary distortions and play no role in the definition of the ferroic properties of a given phase.

However, ferroelectricity can arise in the case of one irreducible OP if $\vec{k} = \delta \vec{a}^*$ is commensurate (C). In this case, the symmetry of the modulated phase will depend on the type of the modulation wavevector (that is, on the odd (even) value of the integers in the fraction describing $\delta$), the CICR of the OP and the global phase $\Phi$. In the case of $\delta = (2l+1)/2m$ and $\Phi = (2l+1)\pi/4m$ ($l$ and $m$ integers), the possible magnetic space groups are $P_a(Pnm2_1)$ and $P_a(Pna2_1)$, if the order parameter has a symmetry $\Delta_1$ or $\Delta_2$ and $\Delta_3$ or $\Delta_4$, respectively. Here, a polarization $P_z$ is allowed. The possibility that a polarization may result from an irreducible magnetic OP and the role played by the global phase $\Phi$ of the OP have been often overlooked. If $\delta = (2l+1)/2m$ and $\Phi = 0$, the phase will be ferroelastic ($e_{xz} \neq 0$), regardless of the symmetry of the OP. Ferrotoroidal moments and the linear magnetoelectric effect can occur only if $\delta = (2l+1)/(2m+1)$ or $\delta = 2l/(2m+1)$ when $\Phi \neq 0$, while ferromagnetism is possible for this type of wavevector only if $\Phi = 0$. These conclusions stress the essential difference between C and INC phases regarding the possible secondary parameters.

*3.2. The case of a reducible and incommensurate OP*
The possible superspace groups obtained for one incommensurate and reducible OP will depend not only on the irreducible representations chosen but also on the phase difference between the irreducible components of the modulation. The different possible symmetries of magnetic phases resulting from the combinations of different pairs of irreducible co-representations show that a polarization can arise in the cases anticipated from the "spiral picture": if the phase difference between the components is $\Delta\Phi = \pi/2$, the magnetic superspace groups allow spontaneous polarizations along the $b$-axis (in the cases $\Delta_1 + \Delta_4$, $P_a(P_{\bar{1}S1}^{n2_1a})$, and $\Delta_2 + \Delta_3$, $P_a(P_{\bar{1}SS}^{n2_1a})$) or along the $c$-axis (in the cases

$\Delta_1 + \Delta_2$, $P_a(P_{\bar{1}11}^{nm2_1})$, and $\Delta_4 + \Delta_3$, $P_a(P_{\bar{1}SS}^{nm2_1})$). In addition, a polarization parallel to the modulation wavevector can arise, if the modulated phase originates from two components of the same symmetry. For example, the OP $G_x + C_y$ (symmetry $\Delta_1 + \Delta_1$) will give rise to a phase with the symmetry $P_a(P_{\bar{1}11}^{2_1ma})$ and, consequently, to a polarization $\vec{P}//\vec{a}$, if $\Delta\Phi \neq 0$.

The generalized time reversal operation $\{\theta;000,1/2\}$, denoted by the prefix $P_a$ in the labeling of the symmetry groups given above, is always a symmetry operation of this type of magnetic phase. Consequently, homogeneous magnetizations, toroidal moments and linear magneto-electric coefficients are forbidden by symmetry. For example, in the case of the cycloidal phase of TbMnO$_3$ (symmetry $\Delta_2 + \Delta_3$, OP $A_x + A_y$, $\Delta\Phi = \pi/2$), the magnetic superspace group is $P_a\left(P_{\bar{1}SS}^{n2_1a}\right)$. This phase is polar along the $b$-axis ($P_y \neq 0$), as observed, but it is neither magneto-electric nor multiferroic. If $\Delta\Phi = 0$, the superspace group would be $P_a(P_{\bar{1}1S}^{11\frac{2_1}{a}})$ and a homogeneous lattice deformation $e_{xy}$ would occur. For a general phase difference between the two components of the order parameter, the superspace group is $P_a(P_{\bar{1}1S}^{11a})$ and, in addition to the secondary modes $P_y$ and $e_{xy}$ previously mentioned, $P_x$ is also possible. Only in the case of a general $\Delta\Phi$, the system is an elasto-electric bi-ferroic, as a ferroeastic deformation and a spontaneous ferroelectric polarization co-exist. However, magneto-electric bi-ferroicity is forbidden.

*3.3-Reducible C- phases: a complex landscape*

The case of a reducible and commensurate phase is more complex because, in this case, the symmetry of the modulated phase depends both on the relative and global phases of the irreducible components of the OP. Let us consider, as illustrative examples, the cases of commensurate phases with $\delta = 1/4$ resulting from modulations with symmetry $\Delta_2 + \Delta_3$ or $\Delta_2 + \Delta_1$. In the first case ($\Delta_2 + \Delta_3$), the possible magnetic space groups are $P_a(P\bar{1})$ if $\Phi_2 = \Phi_3 = 0$, $P_a(P112_1)$ if $\Phi_2 = \Phi_3 = \pi/8$, $P_a(P12_11)$ if $\Phi_{2(3)} = 0, \Phi_{3(2)} = \pi/2$ or $P_a(P1)$, in the other cases. Note that one OP of this symmetry can stabilize phases that are polar either along the $b$-axis ($P_a(P12_11)$) or along the $c$-axis ($P_a(P112_1)$). Note also that, in the latter case, the components of the OP would be in-phase. In the second example ($\Delta_2 + \Delta_1$), the possible magnetic space groups would be $P_a(P1\frac{2_1}{m}1)$ if $\Phi_2 = \Phi_1 = 0$, $P_a(Pnm2_1)$ if $\Phi_{2(1)} = \pi/8, \Phi_{1(2)} = 5\pi/8$ (or equivalent domains) or $P_a(P1m1)$, in the other cases.

In TbMnO$_3$, the phase stable at zero magnetic field below $T_{c2}=28K$ is a $\Delta_2 + \Delta_3$ (mode $A_x + A_y$) incommensurate cycloidal phase with $\Delta\Phi = \pi/2$ [27]. As seen, this phase has symmetry $P_a(P_{\bar{1}SS}^{n2_1a})$ and it is polar along the $b$-axis. Above a critical threshold, an external magnetic field induces a transition to a $\delta = 1/4$ commensurate phase, polar along the $c$-axis. The magnetic structure of this field induced phase has recently been identified as a $\Delta_2 + \Delta_1$ phase (mode $A_x + A_z$), with $\Delta\Phi \approx \pi/2$ [28]. Its symmetry may therefore correspond either to $P_a(P1m1)$ or $P_a(Pnm2_1)$, depending on the global phases adopted. Given that $P$ is parallel to the $c$-axis, the latter case must be chosen. Here, the importance of the global phases is again stressed. In such a case, the field induced transition is $P_a(P_{\bar{1}SS}^{n2_1a}) \rightarrow P_a(Pnm2_1)$, which implies a rotation of the plane of the cycloid (in agreement with the

"spiral picture") and a change of the symmetry of the order parameter, along with a rotation of the polarization from the *b*-axis to the *c*-axis. However, another possibility could be a transition to the phase $P_a(P112_1)$ (a $A_x + A_y$ phase with $\Phi_2 = \Phi_3 = \pi/8$ (or equivalent phases $(2n+1)\pi/8$)), which would also allow for a polarization strictly directed along the *c*-axis, maintaining the symmetry $\Delta_2 + \Delta_3$ of the OP. This possibility, for example, is not commonly taken into account and can not be anticipated from the usual "spiral picture".

**4. Magneto-electric effect and the polarization rotation under field: the example of TbMnO$_3$**

In both the cases considered above, a field driven transition from $P_a(P_{1SS}^{n2_1a})$ to $P_a(Pnm2_1)$ or $P_a(P112_1)$ involves phases with a symmetry that forbids a linear magneto-electric effect. In these circumstances, can the magneto-electric effect play a relevant role in the mechanism for the field driven rotation of the electric polarization?

The transition from the cycloidal INC phase to the cycloidal C phase, observed in TbMnO$_3$ under a magnetic field, is found to be mediated by an intermediate phase [29]. Therefore, in order to answer the question above, one must first clarify the origin of a phase sequence $P_a(P_{1SS}^{n2_1a}) \to ? \to P_a(Pnm2_1)$ and elucidate the nature and the magneto-electric properties of the intermediate phase.

The mechanism for the field induced destabilization of the cycloidal $P_a(P_{1SS}^{n2_1a})$ phase in TbMnO$_3$ has been recently analyzed [30]. The form of the free energy expansion for the case of this incommensurate phase is entirely dictated by the symmetry of the OP:

$$f_{inc} = \frac{\alpha_2}{2}S_2^2 + \frac{\beta_2}{4}S_2^4 + \frac{\alpha_3}{2}S_3^2 + \frac{\beta_3}{4}S_3^4 + \frac{\gamma_1}{2}S_2^2S_3^2\cos(2\varphi) + \frac{\gamma_2}{4}S_2^4S_3^4\cos^2(2\varphi) +$$
$$+ v_1P_yS_2S_3\sin(\varphi) + v_2e_{xy}S_2S_3\cos(\varphi) + v_3P_xS_2^2S_3^2\sin(2\varphi) + \frac{k_y}{2}P_y^2 + \frac{k_x}{2}P_x^2 +$$
$$+ \frac{k_e}{2}e_{xy}^2 + \frac{\mu}{2}M^2 + \frac{\Delta}{2}M^2S_2^2S_3^2\cos(2\varphi) - MB + ... \qquad (1)$$

Here, we have considered the external magnetic field *B* and the induced magnetization *M*, together with a coupling term between *M* and the invariant $S_2^2S_3^2\cos(2\varphi)$, which depends on the relative phase $\Delta\Phi = \varphi$. The equilibrium value of $\varphi$ will depend on the external field and on the relative values of the expansion coefficients. Let $\bar{\gamma}_1 = [\gamma_1 + v_1^2/2k_y - v_2^2/2k_e]$ and $\bar{\gamma}_2 = [\gamma_2 + 2v_3^2/k_x]$. The higher symmetry solutions $\varphi = 0$ or $\varphi = \pi/2$ will be stable if $[\bar{\gamma}_1 + \bar{\gamma}_2S_2^2S_3^2\cos(2\varphi) + \Delta B^2\bar{\mu}^{-2}(\varphi)] > 0$ or $[\bar{\gamma}_1 + \bar{\gamma}_2S_2^2S_3^2\cos(2\varphi) + \Delta B^2\bar{\mu}^{-2}(\varphi)] > 0$, respectively. At zero field, the latter condition is verified experimentally, implying that $[\bar{\gamma}_1 - \bar{\gamma}_2S_2^2S_3^2] > 0$. Also, the destabilization of the cycloidal polar phase by the external magnetic field requires that $\Delta < 0$. Under these conditions, the INC phase is destabilized by a magnetic field $B > B_{c1}$, where $B_{c1}$ is given by:

$$B_{c1} = (\mu + |\Delta|S_2^2S_3^2)\sqrt{\frac{\bar{\gamma}_1 - \bar{\gamma}_2S_2^2S_3^2}{|\Delta|}} \qquad (2)$$

These results have been established in [30]. One can, however, explore a little further this mechanism. Let us first note that the phase $\varphi = 0$ is potentially stabilized for fields $B > B_{c2}$, where:

$$B_{c2} = (\mu - |\Delta| S_2^2 S_3^2) \sqrt{\frac{\bar{\gamma}_1 + \bar{\gamma}_2 S_2^2 S_3^2}{|\Delta|}} \qquad (3)$$

If, for simplicity, we assume that $\mu \gg |\Delta|$, so that $\bar{\mu}(\varphi) \approx \mu$, then $B_{c2}^2 \approx B_{c1}^2 + 2\mu^2 \bar{\gamma}_2 S_2^2 S_3^2 |\Delta|^{-1}$. Since $\bar{\gamma}_2 > 0$, $B_{c2} > B_{c1}$. This means that, for the intermediate field range $B_{c1} < B < B_{c2}$, the phases corresponding to $\varphi = \pi/2$ and $\varphi = 0$ will be both unstable and there must necessarily exist an intermediate phase of lower symmetry. The stability of this intermediate phase requires that:

$$\cos(2\varphi) \approx \frac{1}{\bar{\gamma}_2 S_2^2 S_3^2} \left[ \frac{\lfloor \Delta \rfloor}{\mu^2} B^2 - \bar{\gamma}_1 \right] \qquad (4)$$

If $B=B_{c1}$, this condition imposes $\varphi = \pi/2$. If $B=B_{c2}$, then $\varphi = 0$. That is, above $B_{c1}$, the field drives a continuous variation of the relative phase of the two components of the spin modulation, from $\varphi = \pi/2$ to $\varphi = 0$. This drift of $\varphi$ reduces, as seen, the symmetry of the phase from $P_a(P_{1SS}^{n2_1 a})$ to $P_a(P_{11S}^{11a})$, if we neglect the breaking of $\{\theta;000,1/2\}$ and mirror symmetry induced by the external magnetic field. This reduction of symmetry costs energy and allows the onset of an effective and non-linear magnetoelectric coupling. The increase in energy with respect to the polar cycloidal phase $\varphi = \pi/2$ can be expressed as a function of the magnetic field as:

$$\Delta f_{inc} = f_{inc}(\varphi) - f_{inc}(\frac{\pi}{2}) = \frac{1}{4\bar{\gamma}_2} \left[ \frac{|\Delta|^2}{\mu^4} B^4 - [\bar{\gamma}_1 - \bar{\gamma}_2 S_2^2 S_3^2]^2 \right] = \frac{1}{4\bar{\gamma}_2} \frac{|\Delta|^2}{\mu^4} [B^4 - B_{c1}^4] \qquad (5)$$

Also, in the intermediate field range, the effective magneto-electric and magneto-elastic coupling can be explicitly established from the field dependence of the secondary parameters:

$$P_y = -\frac{\upsilon_1}{k_y} \sqrt{\frac{|\Delta|}{2\bar{\gamma}_2 \mu^2} (B_{c2}^2 - B^2)} \qquad (6a)$$

$$P_x = -\frac{\upsilon_3 |\Delta|}{k_x \bar{\gamma}_2 \mu^2} \sqrt{B^2 (B_{c1}^2 + B_{c2}^2) - B_{c1}^2 B_{c2}^2 - B^4} \qquad (6b)$$

$$e_{xy} = -\frac{\upsilon_2}{k_e} \sqrt{\frac{|\Delta|}{2\bar{\gamma}_2 \mu^2} (B^2 - B_{c1}^2)} \qquad (6c)$$

The increase of the energy of the intermediate $P_a(P_{11S}^{11a})$ phase, given by (5), is driven by the magnetic field via non-linear magneto-electric and magneto-elastic effects. This reduces the relative stability of the phase with respect to potential competing phases. In this process, the magneto-electric coupling induces a rotation of the polarization but solely in the *ab*-plane. The rotation of the polarization from the *ab*-plane to the *c*-axis results from a separate mechanism: the stabilization, via a first order transition, of a different competing (commensurate) phase.

**References**
[1]  Kimura T, Goto T, Shintani H, Ishizaka K, Arima T and Tokura Y 2003 *Nature* **426** 55
[2]  Hur N, Park S, Sharma P A, Ahn J S, Guha S and Cheong S W, 2004 *Nature* **429** 392
[3]  Lawes G, Harris A B, Kimura T, Rogado N, Cava R J, Aharony A, Entin-Wohlman O, Yildirim T, Kenzelmann M, Broholm C and Ramirez A P 2005 *Phys. Rev. Lett.* **95** 087205
[4]  Kimura T, Lashley J C and Ramirez A P 2006 *Phys. Rev. B* **73** 220401(R)
[5]  Yamasaki Y 2006 *Phys. Rev. Lett.* **96** 207204
[6]  Taniguchi K, Abe N, Takenobu T, Iwasa Y and Arima T 2006 *Phys. Rev. Lett.* **97** 097203
[7]  Heyer O, Hollmann N, Klassen I, Jodlank S, Bohatý L, Becker P, Mydosh J A, Lorenz T and Khomskii D 2006 *J. Phys.:Cond. Matter* **18** L471
[8]  Kimura T, Lawes G and Ramirez A P 2005 *Phys. Rev. Lett.* **94** 137201
[9]  Cheong S W and Mostovoy M 2007 *Nature Materials* **6** 13
[10] Teague J R, Gerson R and James W J 1970 *Solid State Commun.* **8** 1073
[11] Bertaut E F, in *Magnetism*, edited by G. T. Rado, H. Suhl (Academic Press, New York, 1963), Vol. III 149
[12] Kimura T, Kawamoto S, Yamada I, Azuma M, Takano M and Tokura Y 2003 *Phys. Rev. B* **67** 180401
[13] Mashiyama H 1980 *J. Phys. Soc. Jpn* **49** 2270
[14] Gesi K, 1986 *Ferroelectrics* **66** 269
[15] Perez-Mato J M 1988 *Solid State Commun.* **67** 1145
[16] Sergienko I A, Sen C, Dagotto E 2006 *Phys. Rev. Lett.* **97** 227204
[17] Sergienko I A, Dagotto E 2006 *Phys. Rev. B*. **73** 094434
[18] Katsura H, Nagaosa N, Balatsky A V 2005 *Phys. Rev. Lett.* **95**, 057205
[19] Mostovoy M 2006 *Phys. Rev. Lett.* **96** 067601
[20] Harris A B 2007 *Phys. Rev. B* **76** 054447
[21] Schweizer J 2005 *C. R. Phys.* **6** 375
[22] Radaelli P G, Chapon L C 2007 *Phys. Rev. B* **76** 054428
[23] Ribeiro J L 2007 *Phys. Rev. B* **76** 144417
[24] Kovalev O V, *Representations of Crystallographic Space Groups: Irreducible Representations, Induced Representations and Corepresentations* 1993 Gordon and Breach, Amesterdam
[25] Pérez-Mato J M, Madariaga G, Tello M J 1984 *Phys. Rev. B* **30** 1534
[26] Here, $P_a(P_{\bar{1}S1}^{nma})$, for example, denotes $P_{\bar{1}S1}^{nma} \times [\{E;000,0\},\{\theta;000,1/2\}]$, where $P_{\bar{1}S1}^{nma}$ is the unitary superspace group.
[27] Kenzelmann M, Harris A B, Jonas S, Broholm C, Schefer J, Kim S B, Zhang C L, Cheong S W, Vajk O P and Lynn J W 2005 *Phys. Rev. Lett.* **95** 087206
[28] Aliouane N, Schmalzl K, Senff D, Maljuk A, Prokes K, Braden M, Agyriuo D N 2009 *Phys. Rev. Lett.* **102** 207205
[29] Aliouane N, Argyriou D N, Strempfer J, Zegkinoglou I, Landsgesell S, von Zimmermann M 2006 *Phys. Rev. B* **73** 020102(R)
[30] Tolédano P 2009 *Phys. Rev. B* **79** 094416